\documentclass[sigconf]{acmart}

\usepackage{booktabs} 
\usepackage{latexsym}

\usepackage{algorithm}
\usepackage{algpseudocode}

\usepackage{tabularx}
\makeatletter
\def\hlinewd#1{%
\noalign{\ifnum0=`}\fi\hrule \@height #1 %
\futurelet\reserved@a\@xhline}
\makeatother

\usepackage{multirow}

\usepackage{array}
\newcolumntype{M}{>{\centering\arraybackslash}m{\dimexpr.1\linewidth-2\tabcolsep}}
\newcolumntype{P}[1]{>{\raggedright\arraybackslash}p{#1}}

\setcitestyle{numbers,sort&compress}

\usepackage{flushend}

\setcopyright{none}

\acmDOI{}

\acmISBN{}

\acmConference[]{}{}{}
\acmYear{}
\copyrightyear{}

\acmArticle{}
\acmPrice{}

\renewcommand\footnotetextcopyrightpermission[1]{}
\begin{document}
\title{Assessing public health interventions using Web content}
\subtitle{Extended Abstract}

\author{Vasileios Lampos}
\orcid{0000-0001-8555-2063}
\affiliation{%
  \institution{University College London}
  \city{London} 
  \country{UK} 
}
\email{v.lampos@ucl.ac.uk}






\begin{abstract}
Public health interventions are a fundamental tool for mitigating the spread of an infectious disease. However, it is not always possible to obtain a conclusive estimate for the impact of an intervention, especially in situations where the effects are fragmented in population parts that are under-represented within traditional public health surveillance schemes. To this end, online user activity can be used as a complementary sensor to establish alternative measures. Here, we provide a summary of our research on formulating statistical frameworks for assessing public health interventions based on data from social media and search engines (Lampos et al., 2015 \cite{lampos2015intervention}; Wagner et al., 2017 \cite{wagner2017intervention}). Our methodology has been applied in two real-world case studies: the 2013/14 and 2014/15 flu vaccination campaigns in England, where school-age children were vaccinated in a number of locations aiming to reduce the overall transmission of the virus. Disease models from online data combined with historical patterns of disease prevalence across different areas allowed us to quantify the impact of the intervention. In addition, a qualitative evaluation of our impact estimates demonstrated that they were in line with independent assessments from public health authorities.
\end{abstract}

%
%



\maketitle

\section{Introduction}
\label{Section:Introduction}
Data generated directly or indirectly by online users ---also simply referred to as user-generated data (UGC)--- can reveal a significant amount of information about their offline behaviour and status. In fact, many recent research efforts have leveraged social media content or search engine usage to address interesting questions in a number of domains, ranging from the Social Sciences \cite{Bakshy2015, Gil2012, Kosinski2013} to Psychology \cite{Kramer2014, manago2012, schwartz2013} and Health \cite{choudhury2013, ginsberg2009, lampos2012}. 

Drawing our focus on health-oriented applications, one of the most prominent research tasks has been the derivation of Web-based syndromic surveillance models for infectious diseases. Modelling influenza-like illness (ILI) rates was the first successful example \cite{culotta2010, ginsberg2009, lampos2010pandemic, polgreen2008}, followed by other conditions \cite{chan2011, Gomide2011, Rohart2016}, including mental health disorders \cite{benton2017, choudhury2013}. Criticisms regarding the accuracy of the original disease models \cite{lazer2014, olson2013} have been resolved in follow-up studies by deploying more elaborate approaches \cite{lamb2013twitter, lampos2015gft, lampos2017www}. One of the key motivations behind all the aforementioned works has been the potential of adopting UGC as a complementary sensor to doctor visits or hospitalisations, which are the main sources of information in traditional public health surveillance networks. An other important factor is that online data could provide access to the bottom of a disease pyramid, i.e. cases of infection present within specific demographies that are not well represented otherwise.

\begin{algorithm*}[t]
    \caption{Assessing the impact of a health intervention using online user-generated data \cite{lampos2015intervention}}
    \label{Alg1}
    \begin{algorithmic}[1]
        \Require{$\mathbf{X}$ (user-generated data), $\mathbf{y}$ (disease rates), $\mathcal{T}$ (target locations where the intervention was applied), $\mathcal{C}$ (control locations; no intervention), $\Delta t_{r}$ (pre-intervention time period), $\Delta t_{\alpha}$ (intervention time period), $\rho_{\text{min}}$ (Pearson correlation threshold)}
        
        \Ensure{$\theta$ (percentage of impact), $\epsilon_\theta$ (confidence intervals), $S_\theta$ (statistical significance)}
        
        \State Train a model $f$ that estimates disease rates from user-generated data during $\Delta t_{r}$, $f$: $\mathbf{X} \rightarrow \mathbf{y}$
        
        \State Derive all location subsets $\mathcal{T}_s$, $\mathcal{C}_s$ of $\mathcal{T}$, $\mathcal{C}$ respectively 
        
        \State Compute disease rates $\mathbf{y}_{\mathcal{T}_s}$, $\mathbf{y}_{\mathcal{C}_s}$ during $\Delta t_{r}$ using $f$
        
        \State Compute all pairwise Pearson correlations, $\mathbf{r}_{{\mathcal{T}_s},{\mathcal{C}_s}}$, between the time series of $\mathbf{y}_{\mathcal{T}_s}$ and $\mathbf{y}_{\mathcal{C}_s}$
        
        \For {all pairs between $\mathcal{T}_s$ and $\mathcal{C}_s$}
            \If {$r_{i,j} \geq \rho_{\text{min}}$} \Comment{$i$, $j$ refer to elements of $\mathcal{T}_s$, $\mathcal{C}_s$ respectively}
            
                \State During $\Delta t_{r}$, train a model $h_{ij}$ that estimates the disease rates of a subset of target locations\par
                \hskip\algorithmicindent from a subset of control locations, $h_{ij}$: $\mathbf{y}_{\mathcal{C}_{sj}} \rightarrow \mathbf{y}_{\mathcal{T}_{si}}$
                
                \State Use $f$ to estimate disease rates in $\mathcal{C}_{sj}$ during $\Delta t_{\alpha}$ based on user-generated data, $\mathbf{y}_c$
                
                \State Use $h_{ij}$ and $\mathbf{y}_c$ to project disease rates in $\mathcal{T}_{si}$ from the ones in $\mathcal{C}_{sj}$ during $\Delta t_{\alpha}$, $\mathbf{y}_\tau^{c}$
                
                \State Use $f$ to estimate disease rates in $\mathcal{T}_{si}$ during $\Delta t_{\alpha}$ based on user-generated data, $\mathbf{y}_\tau$
                
                \State Estimate the impact of the intervention at $\mathcal{T}_{si}$ as $\theta_i = \frac{\mu(\mathbf{y}_\tau) - \mu(\mathbf{y}_\tau^{c})}{\mu(\mathbf{y}_\tau^{c})}$
                
                \State Use bootstrapped impact estimates, $\theta_i^b$, to estimate confidence intervals for $\theta_i$, $\epsilon_{\theta_i}$ ($.025$ and $.975$ quantiles)
                    
                    \If {$|\theta_i| > 2\sigma(\theta_i^b)$}
                        \State Consider the impact estimate $\theta_i$ as statistically significant, $S_{\theta_i} = 1$
                    \Else
                        \State $S_{\theta_i} = 0$
                    \EndIf
                    
            \EndIf
        \EndFor
    \end{algorithmic}
\end{algorithm*}

In this work, we go beyond disease modelling by proposing a statistical framework for assessing the impact of a health intervention (against an infectious disease) based on online information. Public health interventions, such as improved sanitation, immunisation programmes or, simply, the promotion of health literacy, assist in reducing the risk of various infections \cite{Cohen2000interventions,nutbeam2000}. However, the absence of routine evaluation systems for such interventions together with the general deficiencies of the existing disease surveillance schemes (e.g. under-represented parts of the populations), enables only partial assessments, especially in situations where interventions are targeting a seasonal disease that is not characterised by the magnitude of a pandemic.

We evaluate our algorithm against two real-world public health interventions. These are two vaccination campaigns against flu launched in England during 2013/14 (Phase A) and 2014/15 (Phase B). Live attenuated influenza vaccines (LAIV) were administered to school age children in various pilot locations, recognising that children are key factors in the transmission of the influenza virus in the general population \cite{Petrie2013trans}. In Phase A, the vaccine was offered to primary school children (4-11 years) only \cite{pebody2014}, whereas in Phase B it was also offered to children from secondary schools (11-13 years) as well as in an expanded set of locations \cite{pebody2015}.

Data from Microsoft's search engine, Bing, and the microblogging service of Twitter are used as the main observations for the proposed impact assessment framework. We deploy nonlinear supervised learning techniques using composite Gaussian Process kernels to model the time series of text frequencies in relation to disease rates in the population. We then utilise this disease model to uncover linear relationships between the disease rates in areas of interest during a time period prior to the intervention. Finally, we exploit this relationship to estimate a projection of disease rates to affected areas had the intervention not taken place. Our analysis yields interesting results, indicating that the intervention reduced ILI rates by more than $20\%$ in Phase A locations and by approximately $17\%$ in primary school areas in Phase B. Both estimates that are in agreement with independent assessments by Public Health England (PHE) \cite{pebody2014,pebody2015}.\footnote{They are in agreement in principle as direct comparisons are not valid.}

\section{Methods}
\label{Section:Methods}

We briefly describe our approach for modelling disease rates from user-generated text and provide an overview of our statistical framework for assessing the impact of a public health intervention.

The estimation of disease rates from online textual information is formulated as a supervised learning task, $f:$ $\mathbf{X}$$\in$$\mathbb{R}^{n \times m}$ $\rightarrow$ $\mathbf{y}$$\in$$\mathbb{R}^n$, where $\mathbf{X}$ represents the frequency of $m$ textual terms over $n$ time intervals, and $\mathbf{y}$ is the disease rate at the same time intervals (as obtained by a public health authority). Provided that nonlinear models tend to outperform linear ones in text regression tasks \cite{lampos2015gft, lampos2014impact, preotiuc2015plos}, we composed and applied a Gaussian Process (GP) kernel for capturing the structure of our observations. GPs are defined as random variables any finite number of which have a multivariate Gaussian distribution. GP methods aim to learn a function $f$: $\mathbb{R}^m \rightarrow \mathbb{R}$ that is specified through a mean and a covariance (or kernel) function, i.e. $f(\mathbf{x}) \sim \mathcal{GP}(\mu(\mathbf{x}),k(\mathbf{x},\mathbf{x}'))$, where $\mathbf{x}$ and $\mathbf{x}'$ (both $\in \mathbb{R}^m$) denote rows of the input matrix $\mathbf{X}$; for a detailed description of GPs, we refer the reader to \cite{rasmussen2006}. By setting $\mu(\mathbf{x}) = 0$, a common practice in GP modelling, we just learn the hyper-parameters of the kernel. We define the following abstract kernel formulation:
\begin{equation}
k(\mathbf{x},\mathbf{x}') = \left(\sum_{z=1}^{Z} k_{\tau}(\mathbf{g}_{z},\mathbf{g}_{z}')\right) + k_{\nu}(\mathbf{x},\mathbf{x}') \, ,
\end{equation}
where $k_{\tau}$ can be any compatible GP kernel in the literature (we use the Rational Quadratic and the Mat\'{e}rn covariance functions in \cite{lampos2015intervention} and \cite{wagner2017intervention} respectively) that is applied on $Z$ categories (or clusters) of textual features,\footnote{We use $Z=4$ categories of textual features based on the number of tokens (1 to 4).} and $k_{\nu}$ captures noise.

\begin{table*}[t]
\begin{center}
\caption{Impact estimates (disease reduction rates) for super-sets of locations in England participating in vaccination programmes as estimated by online user-generated data. Estimates in bold were assessed as statistically significant.}
\label{table:impact_assessment}
\renewcommand*{\arraystretch}{1.1}
\begin{tabular}{l|c|>{\centering\arraybackslash}m{0.21\textwidth}|>{\centering\arraybackslash}m{0.15\textwidth}|c|>{\centering\arraybackslash}m{0.18\textwidth}}
\hlinewd{1.3pt}
\textbf{Phase} & \textbf{Data Source} & \textbf{Target Locations} ($\mathcal{T}$) & \textbf{Num. of Control Locations} ($\mathcal{C}$) & $r(\mathcal{T},\mathcal{C})$ & \textbf{Disease Reduction Rate \%} ($\theta$) \\
\hlinewd{1.3pt}
\multirow{2}{0.7in}{A (2013/14)} & Twitter & All locations  & $8$ & $.86$ & \textbf{-32.72} $(-47.43, -15.62)$\\\cline{2-6}
                             & Bing    & All locations      & $7$ & $.87$ & \textbf{-21.71} $(-32.12, -9.12)$\\
\hlinewd{1.3pt}   
\multirow{5}{0.7in}{B (2014/15)} & Twitter & All locations                  & $10$ & $.89$ &  $-4.51$ $(-25.72, 22.61)$\\\cline{2-6}
                             & Twitter & Primary school cohort                     & $8$  & $.71$ & \textbf{-16.97} $(-30.09, -2.42)$\\\cline{2-6}
                             & Twitter & Secondary school cohort                  & $7$  & $.83$ & $1.41$ $(-19.40, 28.40)$\\\cline{2-6}
                             & Twitter & Primary \& secondary \newline school cohort       & $7$  & $.84$ & $-0.30$ $(-16.71, 19.36)$\\
\hlinewd{1.3pt}
\end{tabular}
\end{center}
\end{table*}

Our methodology for assessing the intervention's impact, influenced by the work presented in \cite{Lambert2008}, will utilise the above disease rate model. It is presented in detail in Alg.~\ref{Alg1}. Assume that there is a set of target areas $\mathcal{T}$, where the intervention is applied, and a set of control areas $\mathcal{C}$, where the intervention has no effect. We firstly compute disease rate estimates for all areas as well as all possible subsets of them ($\mathcal{T}_s$, $\mathcal{C}_s$) from UGC. Ideally, for a target area we wish to compare the disease rates during (and slightly after) the intervention with disease rates that would have occurred, had the intervention not taken place. Of course, the latter information can only be estimated. Focusing on target-control area pairs with strong linear correlations ($\geq \rho_{\text{min}} = .6$) in historical disease rates prior to the intervention ($\Delta t_r$), we hypothesise that this relationship would have been maintained in the absence of an intervention. Therefore, we can learn a linear model ($h$) that estimates the disease rates in a target area based on the disease rates of a control area with data prior to the intervention. Then, we can use this model to project disease rates in a target area during the intervention period ($\Delta t_{\alpha}$), but had the intervention not taken place. Finally, we can quantify the impact of the intervention by computing the relative percentage of difference ($\theta$) between the actual estimated disease rates (from UGC) and the projected ones. Confidence intervals for $\theta$ can be derived via bootstrap sampling \cite{Efron1994bootstrap}, and in particular by both sampling (with replacement) the linear regression's residuals (from $h$) as well as the input data. Provided that the distribution of the bootstrap estimates is unimodal and symmetric, we assess an outcome as statistically significant, if its absolute value is higher than two standard deviations of the bootstrap estimates.

\section{Results and Discussion}
\label{Section:Results}

We first provide a brief overview of the data sets used in our analysis. We then summarise the outcomes of the intervention's impact assessment in both vaccination campaigns (Phase A and B). Finally, we propose potential directions for future research.

\subsection{Data Sets}
\label{Section:Data}
For the 2013/14 vaccination campaign (Phase A), we considered $7$ target and $12$ control areas (see Table 1 in \cite{lampos2015intervention}). We extracted $308$ million tweets (May, 2011 to April, 2014), $2.2$ million of which contained flu-related $n$-grams.\footnote{We used approximately $200$ $n$-grams, listed in the supplementary material of \cite{lampos2015intervention}.} We additionally obtained search query data (December, 2012 to April, 2014) for a smaller time period due to user privacy regulations, which contained approx. $7.7$ million flu-related queries. As the campaign expanded in 2014/15 to include more locations (Phase B) and different school-age children groups, the number of target locations increased to $17$ ($6$ primary, $7$ secondary, and $4$ primary and secondary school cohorts), and $16$ control areas were deployed (see Table 1 in \cite{wagner2017intervention}). For this period, we extracted $520$ million tweets geolocated in England (August, 2011 to August, 2015). This analysis did not use any search engine data. Historical ILI rates at a national level for England were obtained from the Royal College of General Practitioners, representing the number of ILI cases per $100{,}000$ people from 2011 to 2015. 

\subsection{Intervention Impact Assessment}
\label{Section:Impact}
A GP, as described in Section \ref{Section:Methods}, was used for modelling ILI rates from UGC since it outperformed linear alternatives, namely ridge regression \cite{hoerl1970} and elastic net \cite{zou2005}. Using a 10-fold cross validation, the mean absolute error (MAE) for the Twitter-based model during Phase A was equal to $2.2$ (per $100{,}000$ people) with an average Pearson correlation of $r = .85$, whereas the model used in Phase B (trained and tested on more data) resulted to a MAE of $2.4$ and $r = .84$. The model trained on Bing data (Phase A) outperformed other models on average ($\text{MAE} = 1.6$, $r = .95$), but at the same time was tested on a significantly shorter time span.\footnote{A more detailed performance evaluation is provided in Section 4.1 of \cite{lampos2015intervention}.}

To assess the impact of the LAIV campaign, we first needed to identify control areas with estimated ILI rates that were strongly correlated to rates in the target vaccinated locations before the start of the intervention. In doing so for Phase A (2013/14), we looked for correlated areas in a pre-vaccination period that included the previous flu season only (2012/13). The reason for this was that the strains of influenza virus may vary between distant time periods \cite{Smith2004mapping} and thus, disease rates may be non homogeneous. For Phase B (2014/15), however, we could not anymore use the previous flu season to establish relationships, given that the Phase A campaign had already violated the assumed geographical homogeneity for 2013/14. Thus, we resided to using the period 2011/13\footnote{Includes two flu seasons from August, 2011 to August, 2013.} based on the fact that the circulated flu strains were not characterised by any significant anomalies. Nevertheless, that resulted in less robust estimates as indicated by our bootstrap sampling analysis (which yielded many of them as not statistically significant) and, taking into account the one-year gap between training and applying, perhaps less accurate projections as well.

A summary of the overall impact assessments is provided in Table \ref{table:impact_assessment}, where outcomes in bold are statistically significant. During Phase A, both data sets (Twitter and Bing) point to significant reductions of disease rates, i.e. from $-21.06\%$ (Bing) to $-32.77\%$ (Twitter) on average. A subsequent sensitivity analysis (see Table 4 in \cite{lampos2015intervention}), where more than one control areas were used to project disease rates indicated that results from Twitter were generally more robust, with the overall impact estimate ($-32.77\%$) being the most consistent one. PHE's own impact estimates compared vaccinated to all non vaccinated areas, and ranged from $-66\%$ based on sentinel surveillance ILI data to $-24\%$ using laboratory confirmed influenza hospitalisations. Note though that these numbers represent different levels of severity or sensitivity, and notably none of these computations was statistically significant \cite{pebody2014}. As a further evaluation point, we observed an analogy between the actual level of vaccine uptake and the estimated impact from our end for a number of areas.

In Phase B, our analysis indicated that areas where primary school children were vaccinated benefited the most with an estimated $\theta$ of $-16.97\%$. However, for the current implementation of the secondary school only vaccination programme, there was no clear evidence of any population wide effect. Both these conclusions are in line with findings of previous studies and complement traditional surveillance sources in exhibiting community wide effects of the LAIV pilot campaign \cite{pebody2014,pebody2015}.

\subsection{Future Work}
\label{Section:Future}

Our approach faces common limitations of research efforts based on unstructured user-generated text. Better methods that automate the semantic interpretation of language can be deployed to derive more accurate results. In fact, in follow-up works, we have proposed techniques that are capable of combining the text statistics (e.g. frequency time series) with a word embedding representation \cite{zou2016iid, lampos2017www, mikolov2013word2vec, mikolov2013iclr}. A further, perhaps more significant limitation, is that the entirety of this work relies on the existence of ground truth. Knowing historical disease rates is essential in order to train a disease model from UGC. However, this may not be possible for places with less established healthcare systems or for new infectious diseases. In addition, even when syndromic surveillance can provide estimates for the prevalence of a disease, it is very likely that these will incorporate demographic biases, carrying them over to any supervised model. Thus, there is a necessity to establish unsupervised disease indicators from UGC. This is a harder problem as it will be difficult to evaluate solutions and one will need to account for the specific demographic biases of the online users in order to produce any viable conclusion. Nevertheless, ongoing work will focus on resolving these issues as well as investigating the framework's applicability in assessing different types of a public health intervention.

\begin{acks}
This work presented in this extended abstract has been supported by the grant EP/K031953/1 (EPSRC, ``i-sense'').
\end{acks}

\bibliographystyle{ACM-Reference-Format}
\bibliography{refs} 

\end{document}